%% file: main.tex
\documentclass[a4paper,11pt]{article}
\pdfoutput=1 

\usepackage{jcappub} 

\usepackage{graphicx}  
\usepackage{epstopdf}  
\usepackage{amsmath}   
\usepackage{dcolumn}   
\usepackage{bm}        
\usepackage{amssymb}   
\usepackage{url}       
\hypersetup{pdfencoding=auto, psdextra}
\usepackage{multirow}  
\usepackage{appendix} 
\usepackage{moresize} 

\usepackage{adjustbox}
\usepackage[utf8]{inputenc}
\usepackage[T1]{fontenc}

\graphicspath{{./}{figures/}}

\title{Relic Neutrino Degeneracies and Their Impact on Cosmological Parameters}

\author[a]{Shek Yeung,}
\author[b,1]{King Lau\note{Corresponding author.}}
\author[a]{and M.-C. CHU}

\affiliation[a]{Department of Physics and Institute of Theoretical Physics, The Chinese University of Hong Kong, Shatin, N. T., Hong Kong, China}
\affiliation[b]{School of Physics and Astronomy, University of Minnesota, Minneapolis, Minnesota 55455, USA}

\emailAdd{terryys@link.cuhk.edu.hk}
\emailAdd{kennylau@umn.edu}
\emailAdd{mcchu@phy.cuhk.edu.hk}

\abstract{In the standard $\Lambda$CDM model, neutrinos are treated as radiation, with their masses and possible degeneracy ignored. In this paper, we compute the impact of a finite relic neutrino degeneracy $\xi$ on the CMB angular power spectra, and obtain constraints on $\xi$ using current cosmological data sets. We find that $\xi \approx O(1)$ is still allowed. We also study the correlations between $\xi$, the Hubble parameter $H_0$, and the spectral index $n_s$. Due to these correlations, the CMB constraints on inflation models are loosened when $\xi$ is fitted together with other cosmological parameters, such that some models excluded at 95\% confidence level by standard fittings without $\xi$ could be revived. Furthermore, the tension in CMB and local measurements of $H_0$ is slightly alleviated. Our results suggest that $\xi$ is a non-negligible physical parameter for cosmological analyses.}

\begin{document}
\maketitle
\flushbottom

\section{Introduction}
\label{Introduction}

The Cosmic Microwave Background (CMB) anisotropy measurement is now one of the most powerful cosmological surveys offering valuable insights in Cosmology. It essentially traces the thermal radiation back to the last-scattering surface at red-shift $z\approx1100$, and the CMB anisotropies are presumably seeded by primordial perturbations. The energy densities of the different components and late-time physics such as the reionization evolution \citep{reion1,reion2} are described by a limited number of parameters, which are sufficient for modelling the cosmic history and predicting the angular power spectra of the CMB. Conversely, when high S/N data is available, these parameters can be inferred to high precision. Currently there is a series of experiments yielding high resolution sky maps of CMB temperature and polarization, which provide constraints on cosmological parameters \citep{WMAP9, Planck_2018_VI}, galaxy cluster physics \citep{ACT_SZ, SPT_lensing}, and inflation models \citep{BKI, BKP, BKX}.

Two of the cosmological parameters which have been studied extensively are the tensor-to-scalar ratio $r$ and the spectral index $n_s$. The popular single-field slow-roll inflation models predict the generation of inflationary gravitational waves which can be quantified by a non-zero $r$. They also predict $n_s$ to deviate slightly from 1, with the extent depending on the values of slow-roll parameters $\epsilon$ and $\eta$ \citep{inflation_review}. Moreover, one can rely on the values and constraints on the $r-n_s$ parameter space to discriminate among inflation models \citep{inflation1,inflation2}. Therefore, the systematics and uncertainties of $r$ and $n_s$ have to be handled meticulously. 

The constraints of these cosmological parameters are model dependent. The standard $\Lambda$CDM model assumes three species of massless active neutrinos, which are predicted to have an average number density of $110\,\mathrm{cm^{-3}}$ per species and temperature $T_\nu \approx 1.9 \mathrm{K}$ today. This cosmic neutrino background has not yet been directly detected. Nevertheless, multiple CMB studies have provided evidences of its existence by measuring the effective number of neutrino species $N_\mathrm{eff}$ \citep{WMAP9,Planck_2018_VI,SPT_damping_tail,ACT_parameters}. 

These cosmological neutrinos can be modelled as a non-interacting Fermi gas which obeys the Fermi-Dirac distribution, and their number and energy densities can be calculated accordingly. In standard cosmology textbooks, the neutrino chemical potential $\mu_\nu$ is assumed to be zero in such calculations \citep[for example][]{Dod03}. Nevertheless, in this study we explore the scenarios with a non-zero $\mu_\nu$. If neutrino is Majorana, $\mu_\nu = 0$ \citep{mu0_majorana}. Therefore, if $\mu_\nu \neq 0$ is observed, neutrinos must be Dirac particles, and the measurement of $\mu_\nu$ is of fundamental importance.

In this paper, we study the observational constraints of cosmological neutrino degeneracy $\xi$ (defined in Sec. \ref{xi and CMB}) derived from CMB measurement and cosmological surveys, and the impacts of a non-zero $\xi$ on other cosmological parameters. In Sec. \ref{xi and CMB}, we discuss the implications of $\xi \neq 0$ on the cosmic expansion history, Big Bang Nucleosynthesis (BBN), CMB, and neutrino asymmetry. In Sec. \ref{xi_H0_Degeneracy} and Sec. \ref{xi_ns_Degeneracy}, we reveal the constraining power of CMB data on $\xi$. We also demonstrate the parameter degeneracy between $H_0$ and $\xi$, and that between $n_s$ and $\xi$. In Sec. \ref{Results}, we perform full Markov Chain Monte Carlo (MCMC) fittings for a string of data sets with different combinations of models, and we present the statistical inferences on $\xi$, $n_s$ as well as other cosmological parameters. They are followed by a summary of the significance and implications of our results in Sec. \ref{Conclusions}.

\section{Neutrino Chemical Potential and CMB}
\label{xi and CMB}

Neutrinos are the least understood Standard Model particles. If neutrinos are massless, their cosmological energy density $\rho_\nu$ can be expressed by
\begin{equation}
\label{rho_nu_neff}
\rho_\nu = \rho_\gamma \, \frac{7}{8} \bigg( \frac{4}{11} \bigg)^{4/3} \, N_{\mathrm{eff}},
\end{equation}
where $\rho_\gamma$ is the cosmological photon energy density and $N_{\mathrm{eff}}$ is the effective number of neutrino species, which equals 3.046 for 3 types of active neutrinos \citep{3.046}. However, neutrinos are known to be massive from neutrino oscillation experiments \citep{KamLAND, dayabay}, and they could have non-zero chemical potentials $\mu_{\nu_i}$, where $i = 1, 2, 3$ labels the mass eigenstates. After the neutrino decoupling at temperature $T_{\nu} \approx 1$ MeV, $\mu_{\nu_i}$ enters through a constant parameter $\xi_{\nu_i}=\mu_{\nu_i}/T_{\nu}$, called the neutrino degeneracy factor, in the frozen neutrino distribution function. 

For a single neutrino mass eigenstate $i$ with degeneracy $\xi_{\nu_i}$, its cosmological energy density (at scale factor $a$) is
\begin{equation}
\label{nu energy 1}
\begin{split}
\rho_{\nu_i}(a) & = \frac{1}{(2\pi)^3} \int^{\infty}_{0} \frac{\sqrt{p^2+m_{\nu_i}^2} \, d^3p}{\exp[ \, p/(T_{\nu,0}/a)-\xi_{\nu_i}]+1} \\
& \propto \int^{\infty}_{0} \frac{\sqrt{1+(\frac{am_{\nu_i}}{q})^2} \, q^3 \, dq}{\exp(q-\xi_{\nu_i})+1} \ \bigg( \frac{1}{a} \bigg)^4,
\end{split}
\end{equation}
where $T_{\nu,0}$ is the neutrino temperature today. Note that we have ignored the neutrino mass by setting $E = p$ in the exponential function, as the neutrino decoupling temperature $T_{\nu} \approx $ 1 MeV is much greater than the active neutrino rest mass. The anti-neutrino energy density is simply
\begin{equation} 
\label{nu energy 2}
\rho_{\overline{\nu}_i}(\xi_{\overline{\nu}_i})=\rho_{\nu_i}(-\xi_{\nu_i}).
\end{equation}
Extending to all three mass eigenstates of active neutrinos, we have the total neutrino energy density
\begin{equation} 
\label{nu energy 3}
\rho_\nu = \sum_{i=1}^3 (\rho_{\nu_i}+\rho_{\overline{\nu}_i}). 
\end{equation}

The major impact of $\xi_{\nu_i}$ on CMB physics is that it modifies the expansion rate $H(a)=\dot{a}/a$ through the Friedmann equation
\begin{equation}
\frac{H(a)}{H_0} = \sqrt{\Omega_c a^{-3}+\Omega_b a^{-3}+\Omega_{\gamma}a^{-4}+\Omega_{\Lambda}+\frac{\rho_\nu(a)}{\rho_\mathrm{cr}}} 
\label{nu energy 4}
\end{equation}
in a flat universe, where $H_0 \equiv H(a = 0)$ and $\rho_\mathrm{cr}$ is the critical energy density today. Apart from the expansion rate, $\xi_{\nu_i}$ also affects the evolution of perturbations, since the neutrino energy density and pressure enter the corresponding Boltzmann equations. In this study the neutrino energy density and pressure are modified according to Eq. \eqref{nu energy 1}.

The BBN data provide constraints on the degeneracies of neutrino flavor eigenstates $(\xi_{\nu_e}, \xi_{\nu_\mu}, \xi_{\nu_\tau})$. The value of $\xi_{\nu_e}$ directly affects the neutron-to-proton ratio and hence the primordial $^{4}$He abundance, which is tightly constrained by observation, leading to a small allowed range $-0.008 < \xi_{\nu_e} < 0.013$ \citep{Ioc09}. If there are strong neutrino flavor mixings, $\xi_{\nu_e}$ would be equalized with $\xi_{\nu_\mu}$ and $\xi_{\nu_\tau}$, and the tight bounds from BBN apply to all 3 flavors. Because of the large $\nu_\mu$-- $\nu_\tau$ mixing, $\xi_{\nu_\mu}$ = $\xi_{\nu_\tau}$ is a good approximation \citep{Dol02, Man11}. On the other hand, the recently determined value of $\theta_{13} \approx 8^{\circ}$ \citep{Day15} is small enough to allow for some deviation between $\xi_{\nu_e}$ and $\xi_{\nu_\mu}$. If sterile neutrino exists, the bounds on the neutrino chemical potentials would be further loosened \citep{Dol04}. 

In \cite{Man11}, the BBN constraints on $N_{\mathrm{eff}}$ from primordial neutrino-anti-neutrino asymmetries were calculated with a careful treatment of neutrino interactions and oscillations. The maximum $N_{\mathrm{eff}}$ allowed under the latest Daya Bay best-fit value of $\mathrm{sin}^2(2\theta_{13}) = 0.084$ \citep{Day15} is $3.15$, which translates to $\lvert \xi_{\nu_\mu} \lvert < 0.34$, assuming that $\xi_{\nu_e}$ = 0 and $\xi_{\nu_\mu} = \xi_{\nu_\tau}$. Furthermore, it was shown in \cite{Barenboim} that, once the three-flavor description of the mixed neutrino system is adopted in place of the effective two-flavor description (used in \cite{Man11}), BBN still allows  $\lvert \xi_{\nu_\mu, \nu_\tau} \lvert \approx O(1)$ for suitable values of the initial asymmetries. Such a large value of the neutrino degeneracy may have non-negligible effects on CMB anisotropies, which is the theme of this paper.

The relation between $(\xi_{\nu_1}, \xi_{\nu_2}, \xi_{\nu_3})$ and $(\xi_{\nu_e}, \xi_{\nu_\mu}, \xi_{\nu_\tau})$ can be obtained by considering the neutrino lepton number asymmetry matrix $\mathbf{L}$ and Pontecorvo-Maki-Nakagawa-Sakata (PMNS) Matrix $U_{\mathrm{PMNS}}$ \citep{Barenboim, Barenboim2}, 
\begin{equation}
\begin{split}
U_{\mathrm{PMNS}} = 
&
\begin{pmatrix} 
1 & 0 & 0 \\
0 & c_{23} & s_{23} \\
 0 & -s_{23} & c_{23}
\end{pmatrix} \\
&
\begin{pmatrix} 
c_{13} & 0 & s_{13}e^{-i\delta_{\mathrm{CP}}} \\
0 & 1 & 0 \\
-s_{13}e^{i\delta_{\mathrm{CP}}} & 0 & c_{13}
\end{pmatrix}
\begin{pmatrix} 
c_{12} & s_{12} & 0 \\
-s_{12} & c_{12} & 0  \\
0 & 0 & 1
\end{pmatrix}, \\
\end{split}
\label{PMNS}
\end{equation}
where $c_{ij}/s_{ij}/t_{ij} = \mathrm{cos}\,\theta_{ij}/\mathrm{sin}\,\theta_{ij}/\mathrm{tan}\,\theta_{ij}$, with $\theta_{ij}$ being the neutrino mixing angles, and  $\delta_{\mathrm{CP}}$ is the Dirac CP-violating phase. The neutrino asymmetry matrix in the mass eigenstate basis, $\mathbf{L_m}$, is defined by

\begin{equation}
\begin{split}
\mathbf{L_m} & = \mathrm{diag}(L_1, L_2, L_3), \\
L_i & = \frac{n_{\nu_{i}}-n_{\overline{\nu}_{i}}}{n_{\gamma}},
\label{Lm}
\end{split}
\end{equation}
where $n_{\gamma}$ is the photon number density, and $n_{\nu_{i}}$ is the neutrino number density given by
\begin{equation}
n_{\nu_i} = \frac{1}{(2\pi)^3} \int^{\infty}_{0} \frac{d^3p}{\exp(p/T_{\nu}-\xi_{\nu_i})+1}, 
\end{equation}
for mass eigenstates $i = 1,2,3$. The anti-neutrino number density $n_{\bar \nu_i}$ is given by $n_{\nu_i} (-\xi_{\nu_i})$. The neutrino asymmetry is
\begin{equation}
L_{i} = \frac{1}{12\zeta(3)}\left(\frac{T_\nu}{T_\gamma}\right)^3(\pi^2\xi_{\nu_i}+{\xi_{\nu_i}}^3).
\end{equation}
Similarly one can also define $\mathbf{L_f}$ in the neutrino flavor basis $(\nu_e, \nu_\mu, \nu_\tau)$, and it is related to $\mathbf{L_m}$ by
\begin{equation}
\mathbf{L_m} = U_{\mathrm{PMNS}} \, \mathbf{L_f} \, U_{\mathrm{PMNS}}^{\dagger}.
\label{Lm_Lf}
\end{equation}

In the early universe, the high interaction rate blocks neutrino flavor oscillations and keeps neutrinos in flavor eigenstates. Thus the asymmetry matrix $\bf{L_f}$ is diagonal. However, as shown in \cite{Barenboim}, when the temperature drops below $T_{\nu} = $ 15 MeV, neutrino flavor oscillations become active and off-diagonal components in $\bf{L_f}$ become significant. At around $T_{\nu} \approx 2-5$ MeV, right before BBN and neutrino decoupling, $\bf{L_m}$ becomes diagonal.

\begin{figure}
\centering
\includegraphics[width=9.5cm]{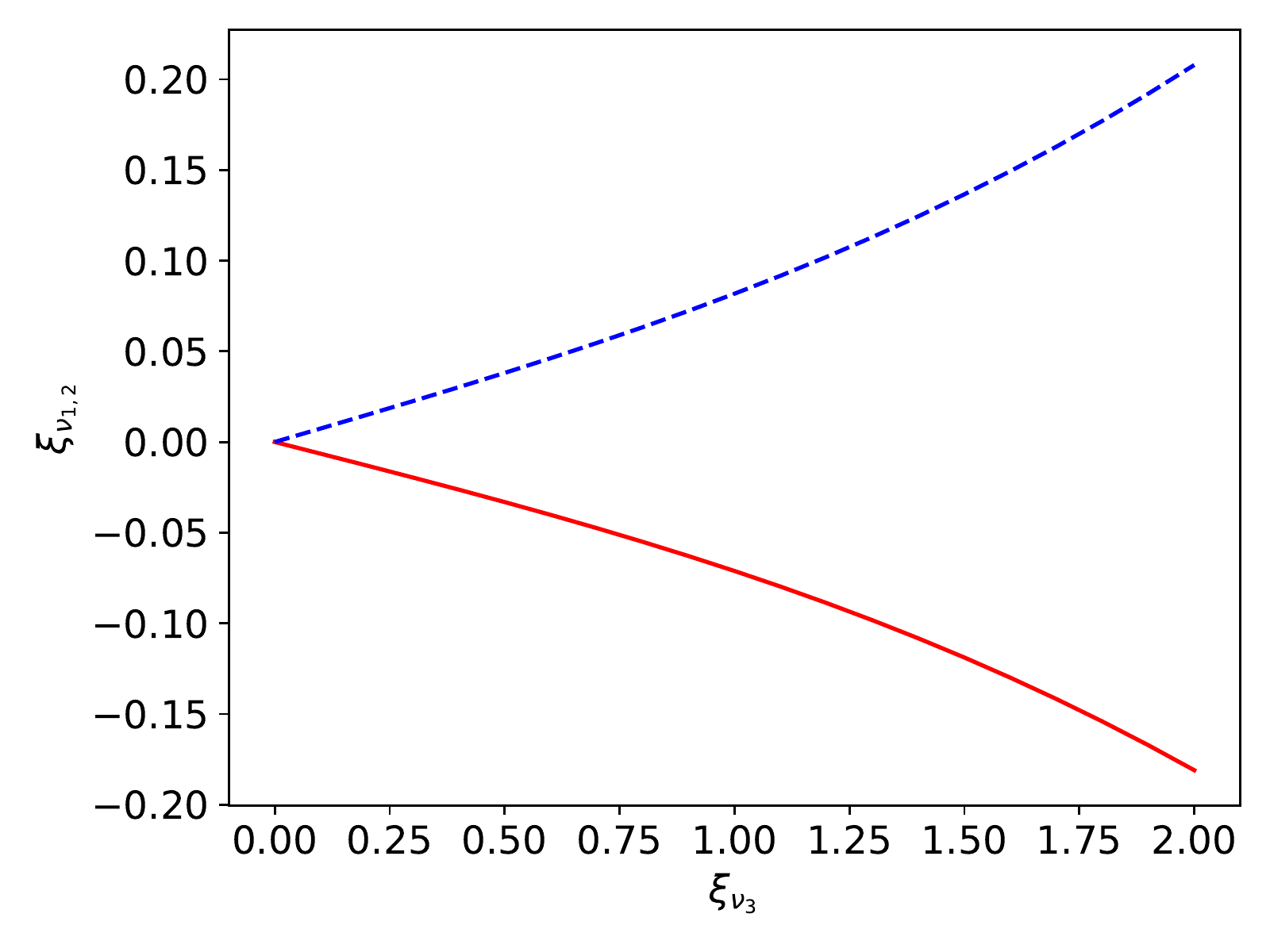}
\caption{Numerical values of $\xi_{\nu_1}$ (red solid line) and $\xi_{\nu_2}$ (blue dashed line) as a function of $\xi_{\nu_3}$. Note that $\xi_{\nu_{1,2}}(-\xi_{\nu_3}) = -\xi_{\nu_{1,2}}(\xi_{\nu_3}) $. A notable feature is $\xi _{\nu_2} \approx -\xi _{\nu_1} \ll \xi_{\nu_3}$ for $\xi_{\nu_3} \in [0, 2]$.}
\label{xi12_depedence}
\end{figure}

By assuming $\delta_{\mathrm{CP}}=0$ and $L_e = 0$, we can rewrite Eq. \eqref{Lm_Lf} to
\begin{align}
L_{e}   & = c_{13}^2(c_{12}^2L_{1}+s_{12}^2L_{2})+s_{13}^2L_{3} = 0, \label{Le}\\
L_{\mu} & = c_{23}[(1-t_{12}^2)c_{23}-2s_{13}s_{23}t_{12}]L_2 \nonumber \\
        & + [(1-t_{13}^2)s_{23}^2-t_{12}t_{13}^2c_{23}(2s_{13}s_{23}+t_{12}c_{23})]L_3, 
        \label{Lmu} \\
L_{\tau} & = s_{23}[(1-t_{12}^2)s_{23}+2s_{13}c_{23}t_{12}]L_2 \nonumber \\
         & + [(1-t_{13}^2)c_{23}^2+t_{12}t_{13}^2s_{23}(2s_{13}c_{23}-t_{12}s_{23})]L_3 
         \label{Lt}.
\end{align}

As $\xi_{\nu_{\mu}}=\xi_{\nu_{\tau}}$, or $L_{\mu}=L_{\tau}$, we can reduce the parameters $(\xi_{\nu_1}, \xi_{\nu_2}, \xi_{\nu_3})$ to one degree of freedom, which we choose to be $\xi \equiv |\xi_{\nu_3}|$. The dependences of $\xi_{\nu_1}$ and $\xi_{\nu_2}$ on $\xi$ are shown in Fig. \ref{xi12_depedence}. In this computation, the values of $c_{ij}/s_{ij}/t_{ij}$ are taken from \cite{pdg_2016}. We can see that $|\xi_1|\approx|\xi_2|\ll\xi$, and therefore the total neutrino asymmetry $L \equiv \sum_{i=1}^3 L_i\approx L_3$. We also take $m_{\nu} = \sum_{\nu_i} m_{\nu_i}/3 \equiv m /3$ in Eq. \eqref{nu energy 1} for all subsequent neutrino energy density calculations. This is to assume that there are three species of degenerate massive neutrinos, following the treatment in \cite{Planck_2018_VI}.

Note that in our derivation of Eq. \eqref{Le}-\eqref{Lt}, we assume $\delta_{CP}=0$. We have calculated an alternative scenario $\delta_{CP}=-\pi/2$, and we found only insignificant changes of the cosmological parameters relative to the case $\delta_{CP}=0$.

A large $\xi$ implies large neutrino asymmetry, which contradicts popular leptogenesis scenarios in which sphalerons effectively transfer lepton asymmetry to baryons in the early universe \citep{Fuk86}. Observational information on $\xi$ therefore has important implications on theories of matter-anti-matter asymmetry in the universe \citep{Man11}, though care has to be taken in relating the lepton asymmetries before and after flavor oscillations \cite{Man11}. On the other hand, a finite neutrino chemical potential would affect the neutrino energy density, which in turn would modify the CMB anisotropy spectra.

The method applied in \cite{Orito} and \cite{Barenboim} calculates the impacts from $\xi$ under the framework of $N_{\mathrm{eff}}$. They use the relation 
\begin{equation}
    \Delta N_{\mathrm{eff}} = \sum_{k} \frac{30}{7}\left(\frac{\xi_{\nu_k}}{\pi}\right)^{2} + \frac{15}{7}\left(\frac{\xi_{\nu_k}}{\pi}\right)^{4},
    \label{eq:neff_massless}
\end{equation} where $k$ labels the mass eigenstates \citep{Dol02}. However, this equation is only true when $m_{\nu}=0$. In the massive neutrino scenario, neutrinos become non-relativistic at low red shift, and hence Eq. \eqref{rho_nu_neff} is no longer valid. That is why we choose to compute neutrino energy density integrals instead of directly adopting such $\Delta N_{\mathrm{eff}}$ to account for $\xi$ in CMB data-fitting. This is also the approach used in some previous studies, for example \cite{Lesgourgues1999, Lattanzi2005, Castorina2012}. In addition, we compare the effects of these two parameters in cosmological data fittings by running two sets of models, one uses $\xi$ as an extra parameter while another uses $N_{\mathrm{eff}}$. We will show they yield two different sets of constraints in Section \ref{section:xi_vs_neff}.

\begin{figure}
\centering
\includegraphics[width=10cm]{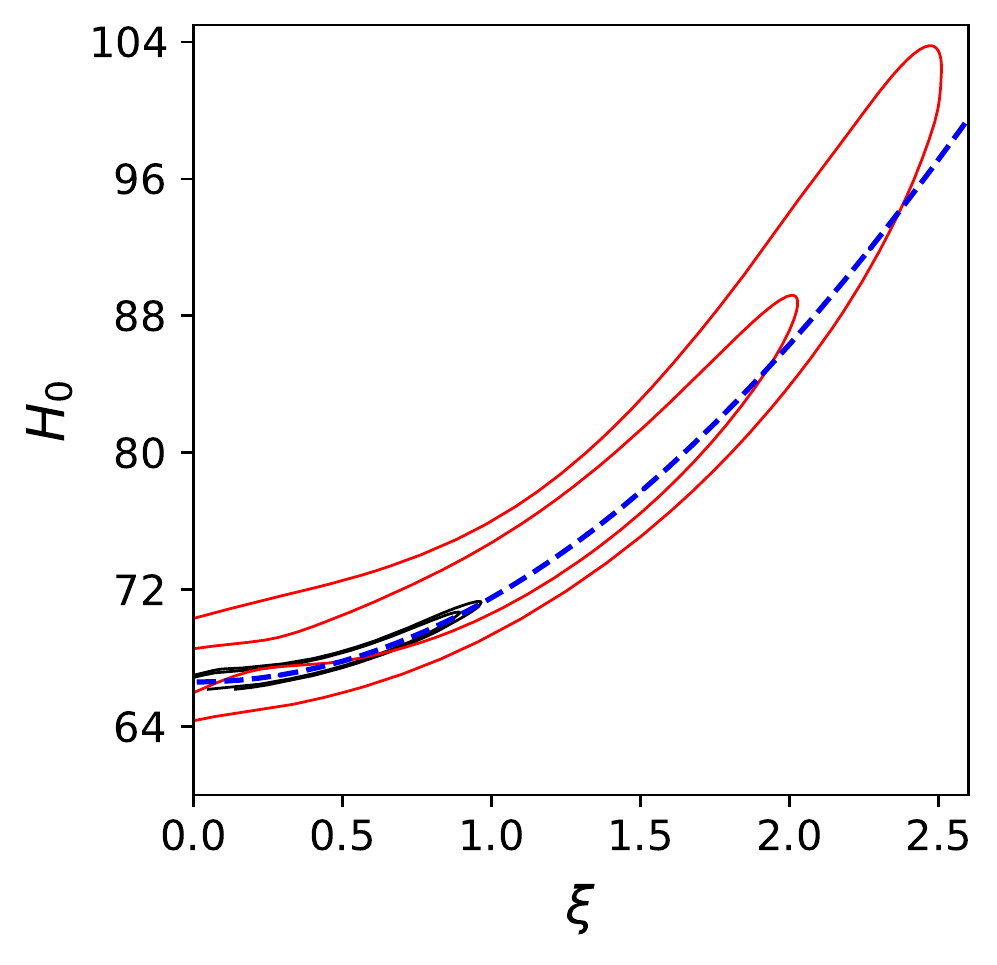}
\caption{Degeneracy between $\xi$ and $H_0$ in CMB fitting. The blue dashed line is the result coming from  the semi-analytical method described in Sec. \ref{xi_H0_Degeneracy}. Two sets of background contours (for 68\% CL and 95\% CL) with different colours are both obtained by MCMC fitting of the \textit{Planck} temperature data, but the black contours is the result from three varying parameters $(\theta_s, \xi)$, while the red contours are from $\Lambda\mathrm{CDM}+\xi$ model.}
\label{xi-H0_correlation}
\end{figure}

\begin{figure}
\centering
\includegraphics[width=13cm]{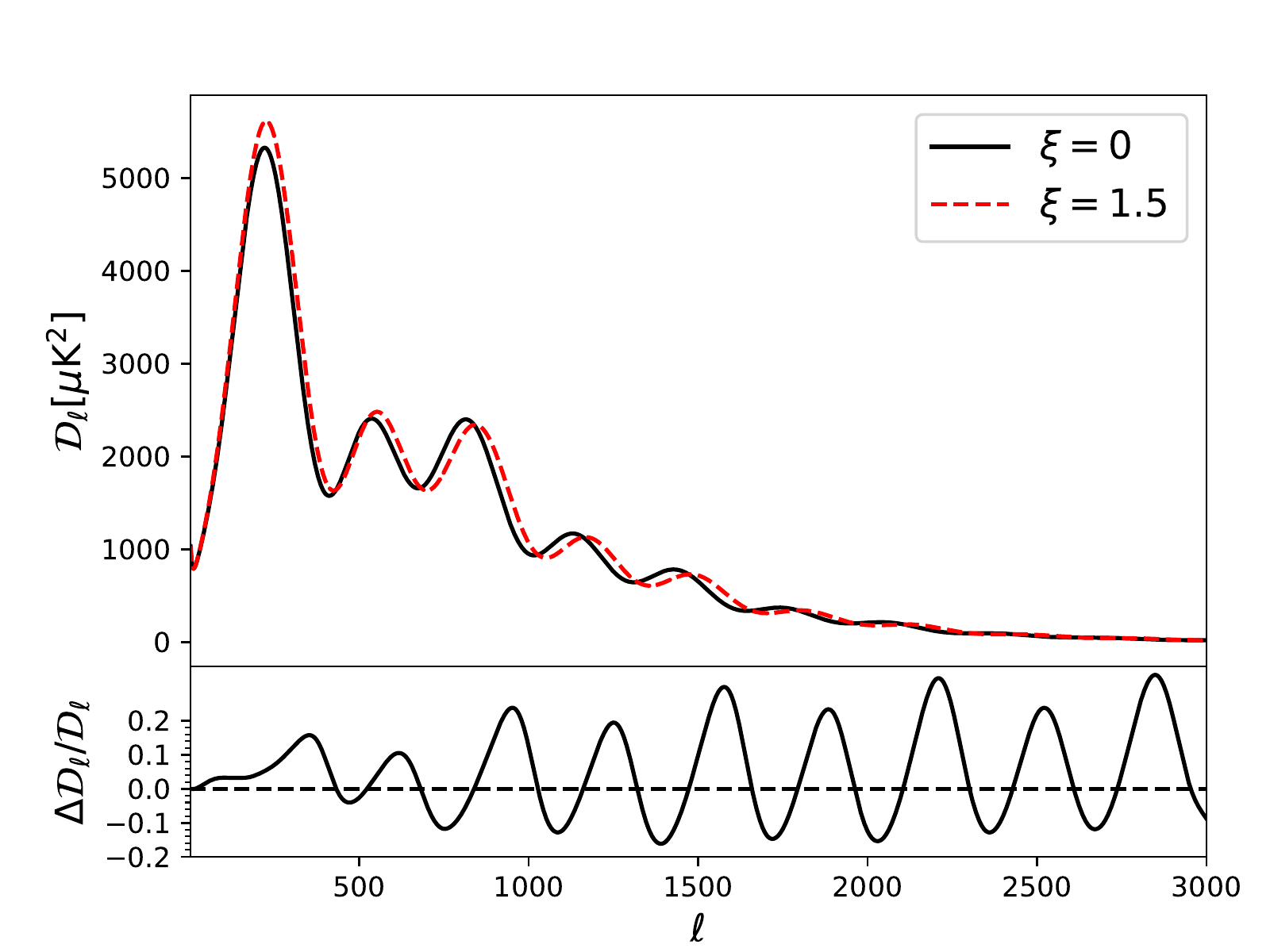}
\caption{Top panel: CMB temperature power spectrum $D_{\ell}^{TT} = \ell(\ell+1)C_{\ell}^{TT}/2\pi$ for $\xi=0$ (black solid line) vs. $\xi=1.5$ (red dashed line). Bottom panel: Fractional changes of the latter compared to the former. It can be seen that a finite $\xi$ shifts the peaks and troughs of the CMB temperature power spectrum. All other cosmological parameters are fixed as best-fits from CMB data.}
\label{cl_xi}
\end{figure}

\begin{figure}
\centering
\includegraphics[width=13cm]{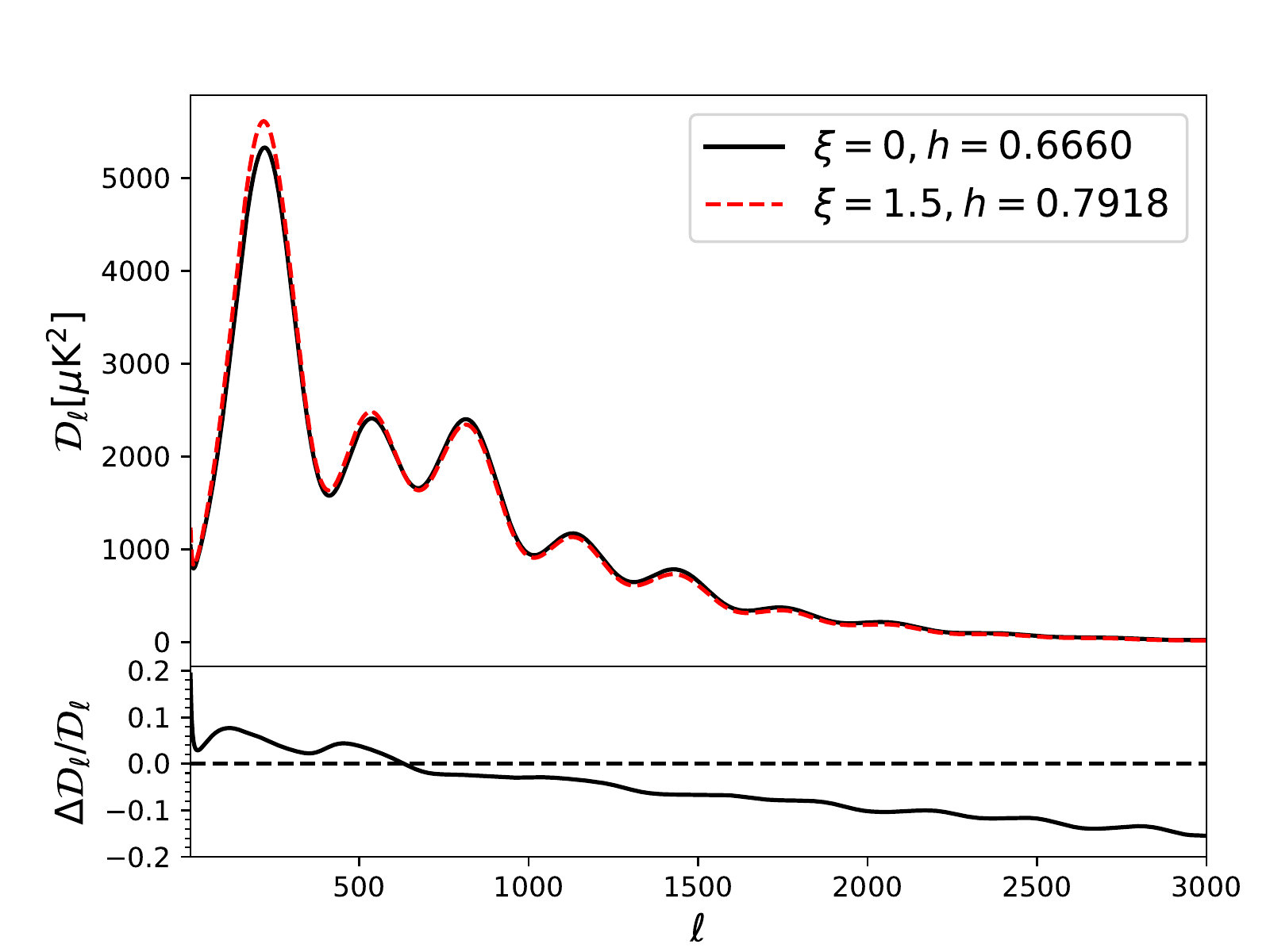}
\caption{CMB temperature power spectrum $D_{\ell}^{TT}$ generated in the same manner as in Fig. \ref{cl_xi}, except that $h$ is also varied to give a fixed $\theta_s$. In the bottom panel, the fractional change appears as a power spectrum tilt.}
\label{cl_xi_h}
\end{figure}

\section{Parameter Degeneracy between $\xi$ and $H_0$}
\label{xi_H0_Degeneracy}

If there is a strong correlation (parameter degeneracy) between $\xi$ and $H_0$ in CMB data fitting, the assumption $\xi=0$ will generate a bias in $H_0$. We change the neutrino energy density according to Eq. \eqref{nu energy 1}-\eqref{nu energy 3} in the CMB power spectrum code \texttt{CAMB}\footnote{homepage: \url{https://camb.info/}} \citep{CAMB1, CAMB2} and perform ``parameter fitting'' (see Sec. \ref{Results} for detail) for the \textit{Planck} CMB temperature data to get constraints on the $\xi-H_0$ parameter space. The result is presented as the contours in Fig. \ref{xi-H0_correlation}, which indeed shows a clear and significant parameter degeneracy between $\xi$ and $H_0$. 

In order to \textit{understand} the origin of this parameter degeneracy, we further explore the underlying CMB physics. The constraining power of CMB temperature power spectrum $C_{\ell}^{TT}$, originates from high precision measurement of the ratios of even to odd peak heights, the amplitudes of the acoustics peaks and the angular scales of peaks and troughs. As a result, tight constraints on the baryon energy density $\rho_b$, the total matter energy density $\rho_m$, and the angular size of the sound horizon at decoupling $\theta_s$ are obtained respectively \citep{Zhen_Hou}. In particular, $\theta_s$ can be calculated by
\begin{equation}
\begin{split}
\theta_s & = \frac{r_s}{D_A} \\
& =  \int_{0}^{a_*} \frac{da}{\sqrt{3(1+R)} \, H(a) a^2} \Bigg/ \int_{a_*}^{1} \frac{da}{H(a) a^2},
\label{theta eq} 
\end{split}
\end{equation}
where $r_s$ is the co-moving radial size of the sound horizon at decoupling, $D_A$ is the co-moving distance of the last scattering surface from us, $a_{*}$ is the scale factor of photon decoupling and $R=3\rho_b/4\rho_{\gamma}$. As $H(a)$ evolves according to Eq. \eqref{nu energy 4} and depends on $\Omega_b$, $\Omega_c$, $\Omega_{\Lambda}$, $H_0$, $\xi$, and $m$, the dependence of $\theta_s$ can be expressed by
\begin{equation}
\theta_s=\theta_s(\Omega_b, \Omega_c, \Omega_{\Lambda}, H_0, \xi, m).
\label{theta dependence}
\end{equation}
This implies that a finite $\xi$ can shift the peak and trough locations of $C_{\ell}^{TT}$, as demonstrated clearly in Fig. \ref{cl_xi}.

In light of the discussion above, we study the $\xi-H_0$ degeneracy semi-analytically. We simply assume that in a typical MCMC CMB data fitting, the parameters are varied so that they give a fixed $\theta_s$. We also assume that $(\Omega_bh^2, \Omega_ch^2, m)$ are essentially fixed during this process, where $h \equiv H_0/100 {\rm km\, s^{-1}\, Mpc^{-1}}$. Eq. (\ref{theta dependence}) is then reduced to $\theta_s=\theta_s(H_0, \xi)$. So we start from evaluating $d\theta_s=0$ through Taylor expansion of Eq. \eqref{theta eq} around $\xi = 0$. It yields
\begin{equation}
c_1 (H_0 \, dH_0) = c_2 \big( \xi \, d \, \xi \big), 
\end{equation}
where
\begin{equation}
c_1 = \frac{3\pi}{4G{T_{\nu,0}}^2} \Bigg(  \int_{0}^{a_*} \frac{c_3(a^{\prime}) \, da^{\prime}}{ {a^{\prime}}^2 \, {H(a^{\prime})}^3}  
- \theta_s \int_{a_*}^{1} \frac{da^{\prime}}{ {a^{\prime}}^2 \, {H(a^{\prime})}^3} \Bigg) ,
\end{equation}
\begin{equation}
\begin{split}
c_2 = & \, \theta_s \int_{a_*}^{1} 
c_4(a^{\prime}) \, \frac{1-{a^{\prime}}^2}{ {a^{\prime}}^4 \, {H(a^{\prime})}^3} \, da^{\prime} \\
& - \int_{0}^{a_*} c_3(a^{\prime}) \, c_4(a^{\prime}) \,
\frac{1-{a^{\prime}}^2}{ {a^{\prime}}^4 \, {H(a^{\prime})}^3} \, da^{\prime},
\end{split}
\end{equation}
\begin{equation}
c_3(a^{\prime}) = \frac{1}{\sqrt{3(1+R(a^{\prime}))}},
\end{equation}
and
\begin{equation}
c_4(a^{\prime}) = \int_{0}^{\infty} \, \frac{6{p^{\prime}}^4+9{p^{\prime}}^2 {m_{\nu_3}}^2+2{m_{\nu_3}}^4 }{(e^{\frac{p^{\prime}a^{\prime}}{T_{\nu,0}}}+1) ({p^{\prime}}^2+{m_{\nu_3}}^2)^{\frac{3}{2}}} \, dp^{\prime}, 
\end{equation}
which reveals that $\xi$ and $H_0$ are clearly correlated. In this calculation, we assume 
$\xi_{\nu_{1,2}} = 0$ (justified by Fig. \ref{xi12_depedence}), $m_{\nu_3}=0.02$ eV, and $(d\Omega_bh^2, d\Omega_ch^2) = 0$. Again, the coefficients $c_1$, $c_2$, $c_3$, and $c_4$ can be calculated numerically using the best-fit parameters from CMB data. 

The results from this method are shown in Fig. \ref{xi-H0_correlation}. Obviously our semi-analytical calculation describes this $\xi-H_0$ correlation very well. When there are only two varying parameters $(\theta_s, \xi)$ in the MCMC fitting, the dashed line aligns almost perfectly with the black contours. In the seven-parameter $\Lambda\mathrm{CDM}+\xi$ fitting, there are additional degeneracies coming from other parameters, but the dashed line still aligns with the general trend of the red contours. Therefore, we conclude that the degeneracy between $\xi$ and $H_0$ in CMB data fitting is due to the fact that both contribute to $\theta_s$, which is tightly constrained by the CMB temperature spectrum.

\section{Parameter Degeneracy between $\xi$ and \texorpdfstring{$\MakeLowercase{n_s}$}{ns}}
\label{xi_ns_Degeneracy}

We use a similar method as in Sec. \ref{xi_H0_Degeneracy} to study the possible degeneracy between $\xi$ and $n_s$. In Fig. \ref{cl_xi_h}, we show two sets of $D_{\ell}^{TT}$, corresponding to $\xi=0$ and $\xi=1.5$, and with $h$ tuned to yield the same $\theta_s$ as the best-fit value from CMB data. All other cosmological parameters in these two cases are also fixed to these best-fit values as well. As expected, the peaks and troughs of $D_{\ell}^{TT}$ are aligned between the two parameter sets. Nevertheless, there is a clear tilting between these two sets of $D_{\ell}^{TT}$ shown in the lower panel of Fig. \ref{cl_xi_h}, which implies that a positive correlation between $\xi$ and $n_s$ would emerge in the CMB data-fitting.

This tilting in the CMB power spectrum can also be caused by a change in $r_s$ and the mean squared diffusion distance $r_D$. When we compare the two ($\xi$, $h$) cases described above, we find that $r_s$ and $r_D$ of ($\xi=1.5$, $h=0.7918$) are about $1.78\%$ and $0.77\%$ smaller than those of ($\xi=0$, $h=0.6660$) respectively, so that the change of $r_D/r_s \approx +1\%$. As the damping of $C_{\ell}^{TT}$ is proportional to $\mathrm{exp}[{(-k/k_D)}^2]$ in $k$-space \citep{damping}, and $k/k_D$ is proportional to $r_D/r_s$, there is a decrease of the CMB temperature power in high $\ell$'s. On the other hand, the increase in CMB low-$\ell$ power, characterized by the growth of the amplitude of the first acoustic peak, is due to the delay of the epoch of matter-radiation equality by $\xi$.

In order to understand quantitatively the origin of $\xi-n_s$ correlation in the CMB constraint, we again use a semi-analytical method. We assume that the impact of $n_s$ can be modelled by $C_{\ell}^{TT} \propto ({\ell}/{\ell}_p)^{n_s-1}$, where ${\ell}_p$ is the pivot scale. Then when there is a change in $n_s$ and other parameters held constant, the ratio between the new and old spectra is
\begin{equation}
\frac{C_{\ell}^{TT}({n_s}^{\prime},h,\xi)}{C_{\ell}^{TT}(n_s,h,\xi)} = \bigg(\frac{\ell}{{\ell}_p(h,\xi)}\bigg)^{{n_s}^{\prime}-n_s}.
\end{equation}
As $\xi \rightarrow \xi^{\prime}$, we look for parameters $({n_s}^{\prime},h^{\prime})$ such that $C_{\ell}^{TT}({n_s}^{\prime},h^{\prime},\xi^{\prime})/C_{\ell}^{TT}(n_s,h,\xi) = 1$ and $\theta_s(h^{\prime},\xi^{\prime})/\theta_s(h,\xi) = 1$. Thus the shift in $n_s$ induced by that of $\xi$ can be described by
\begin{equation}
\label{delta_ns}
\Delta n_s = {n_s}^{\prime}-n_s = \frac{\mathrm{ln}[C_{\ell}^{TT}(n_s,h,\xi)/C_{\ell}^{TT}(n_s,h^{\prime},\xi^{\prime})]}{\mathrm{ln}[{\ell}/{\ell}_p(h^{\prime},\xi^{\prime})]}, 
\end{equation}
and we average over $2 \leqslant {\ell} \leqslant 2508$ to get a single $\Delta n_s$. 

\begin{figure}
\centering
\includegraphics[width=10cm]{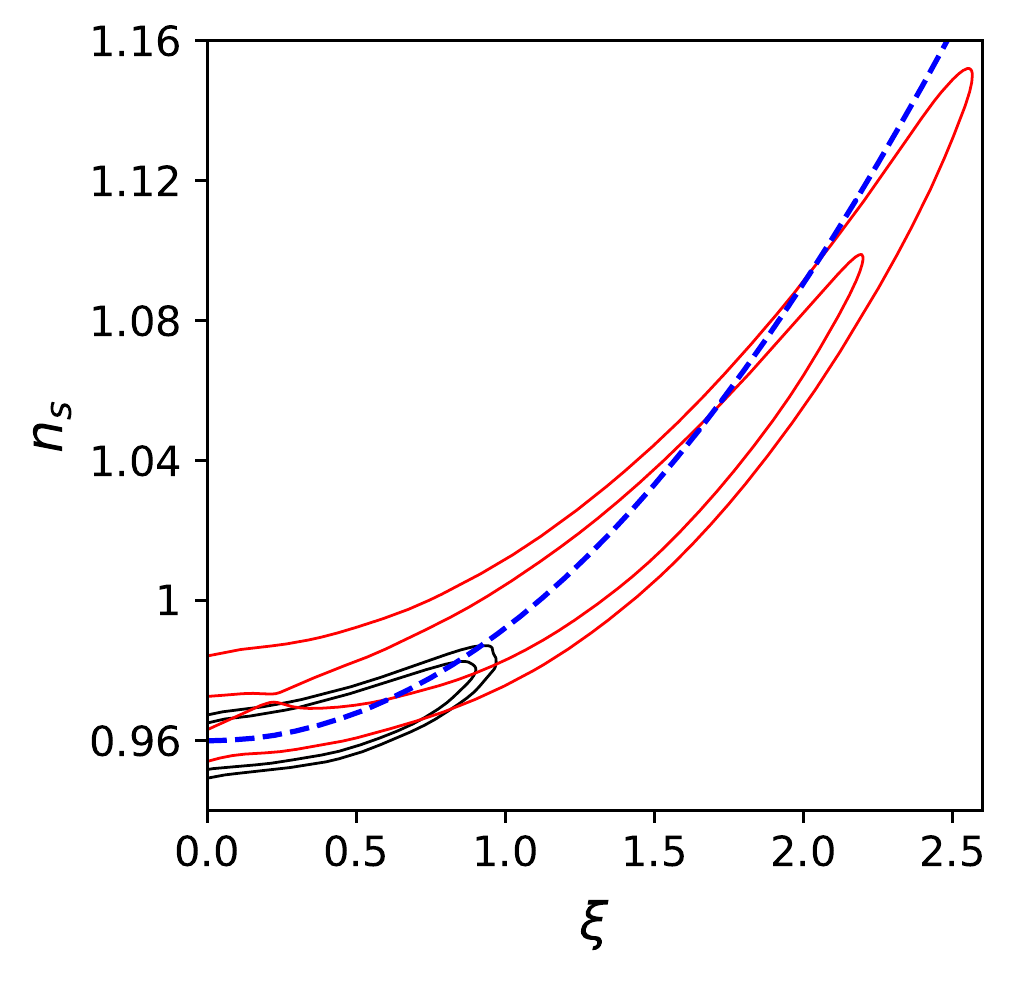}
\caption{Correlation between $\xi$ and $n_s$ by the semi-analytical calculation (blue dashed line). The black contours are obtained by MCMC fitting of the \textit{Planck} temperature data by varying $(\theta_s, n_s, \xi)$, while the red ones are from the seven-parameter fitting of the $\Lambda\mathrm{CDM}+\xi$ model.}
\label{xi-n_s_correlation}
\end{figure}

The contours of constraints imposed by the \textit{Planck} temperature data and result from our semi-analytical method Eq. \eqref{delta_ns} are plotted together in Fig. \ref{xi-n_s_correlation}. The contours indeed show a positive correlation between $\xi$ and $n_s$, and again our semi-analytical method describes this correlation correctly, especially when there are only three varying parameters.

So far we have only considered the constraints imposed by the temperature data. In the next section, we examine whether the same conclusion still holds when various cosmological data sets and models are involved, and investigate the extent to which $\xi$ can affect the constraints of $n_s$.

\section{Fitting Results}
\label{Results}

Here we utilize the 2019 July version of the \texttt{CosmoMC}\footnote{homepage: \url{https://cosmologist.info/cosmomc/}} \citep{COSMOMC1, COSMOMC2} code to apply the MCMC fitting strategy which is widely-adopted in the CMB community. By searching in the multi-dimensional cosmological parameter space for the best agreement between the generated CMB power spectrum and data, \texttt{CosmoMC} finds the posterior probability density functions (as well as their means and standard deviations) for both independent and dependent parameters by likelihood functions based on the cosmological data sets provided.

\begin{table*}[t]
\centering
\begin{adjustbox}{max width=\textwidth}
\begin{tabular}{*{12}{c}}
\hline
\hline 
Model & $\Delta\mathrm{ln}\mathcal{L}$ & $\sigma_h$ & $H_0$ & 100$\Omega_{b}h^2$ & $\Omega_{c}h^2$ & $100\theta_s$ & $\tau$ & $\mathrm{ln}(10^{10}A_s)$ & $n_s$ & $\xi$ & $m$\\
\hline \\
\input{table_data_1_new.txt}

\hline
\hline
\end{tabular}
\end{adjustbox}
\caption {Parameter constraints (mean and 68\% CL values), changes of the best-fit likelihood $\Delta\mathrm{ln}\mathcal{L}$ and the tension $\sigma_h$ (in $\sigma$ unit) from two models. The data sets used are \textit{Planck}(T+P), BAO, DES, and lensing. $L$ represents the $\Lambda\mathrm{CDM}$ case, which is the base-line model with 6 parameters $(\Omega_{b}h^2, \Omega_{c}h^2, 100\theta_s, \tau, \mathrm{ln}(10^{10}A_s), n_s)$. As defined in the text, $\xi = |\xi_{\nu_3}|$ and $m$ is the sum of neutrino masses.}
\label{stat table 1}
\end{table*}

\subsection{Models}

We define our models as follow: the baseline $\Lambda\mathrm{CDM}$ model has 6 major cosmological parameters as the independent input: $(\Omega_{b}h^2, \Omega_{c}h^2, 100\theta_s, \tau, \mathrm{ln}(10^{10}A_s), n_s)$, where $\tau$ is the reionization optical depth, $A_s$ is the power spectrum amplitude and $n_s$ is the spectral index. We also launch a 7-parameter model $\Lambda\mathrm{CDM}+\xi$, as well as a 8-parameter model $\Lambda\mathrm{CDM}+\xi+m$, by implementing the changes to \texttt{CAMB} described in Sec. \ref{xi and CMB}. Similarly, one can replace $\xi$ by $N_\mathrm{eff}$ to get other two models. The default values $N_{\mathrm{eff}}=3.046$ and $m=0.06$ eV in \texttt{CosmoMC} are fixed when they are not varying parameters in data fitting. We use the prior $N_\mathrm{eff}>3.046$ when $N_\mathrm{eff}$ is varied.

\subsection{Data Sets}

Our primary data set is ``\textit{Planck}(T)'' containing CMB temperature power spectrum from the \textit{Planck} Collaboration, while ``\textit{Planck}(T+P)'' includes the additional CMB low-$\ell$ ($2 \leqslant {\ell} \leqslant 29$) and high-$\ell$ ($30 \leqslant {\ell} \leqslant 1996$) polarization spectra released in 2019 \citep{Planck_2018_like}. The polarization data carry independent information and provide extra constraining power, particularly on $\tau$. The data set ``BAO'' which contains data of Baryonic Acoustic Oscillation is from BOSS DR12 ``LOWZ'' and ``CMASS'' surveys \citep{BAO}, and the ``DES'' and ``lensing'' are from Dark Energy Survey lensing data \citep{DES} and \textit{Planck} CMB lensing data \citep{Planck_2018_lensing} respectively. Since they are closely related to the matter power spectrum, we expect that they provide supplementary information about the total neutrino mass. (Note that as \texttt{CosmoMC} relies on \texttt{CAMB} for the calculation of cosmological physics, BAO and lensing data sets are thus also subjected to the changes that we have introduced to the $\xi$ models.)

We have run \texttt{CosmoMC} fittings on all combinations of models with data sets. However, in the following subsections we only discuss the results from the combinations of models $\Lambda\mathrm{CDM}$ and $\Lambda\mathrm{CDM}+\xi+m$, fitted to the data set ``\textit{Planck}(T+P)+BAO+DES+lensing''. All other results and relevant discussions can be found in the Appendix. 

\begin{figure}
\centering
\includegraphics[width=10cm]{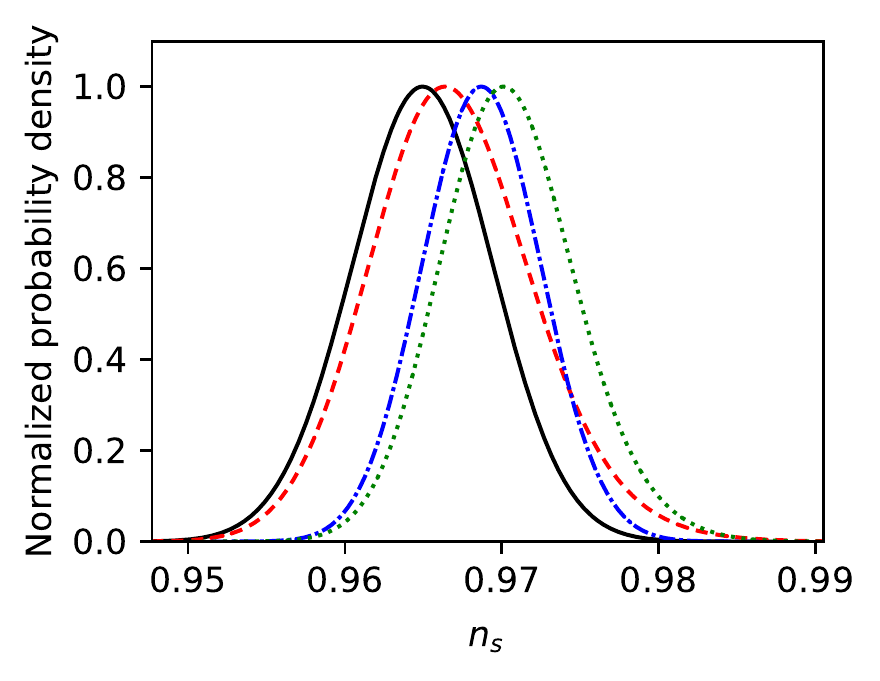}
\caption{1D marginalized pdfs of $n_s$ for \textit{Planck}(T+P) with $\Lambda\mathrm{CDM}$ (black solid), \textit{Planck}(T+P) with $\Lambda\mathrm{CDM}+\xi+m$ (red dashed), \textit{Planck}(T+P)+BAO+DES+lensing with $\Lambda\mathrm{CDM}$ (blue dashed-dotted), and \textit{Planck}(T+P)+BAO+DES+lensing with $\Lambda\mathrm{CDM}+\xi+m$ (green dotted). These distribution functions are normalized with respect to their peak values.}
\label{ns_1D}
\end{figure}

\begin{figure}
\centering
\includegraphics[width=10cm]{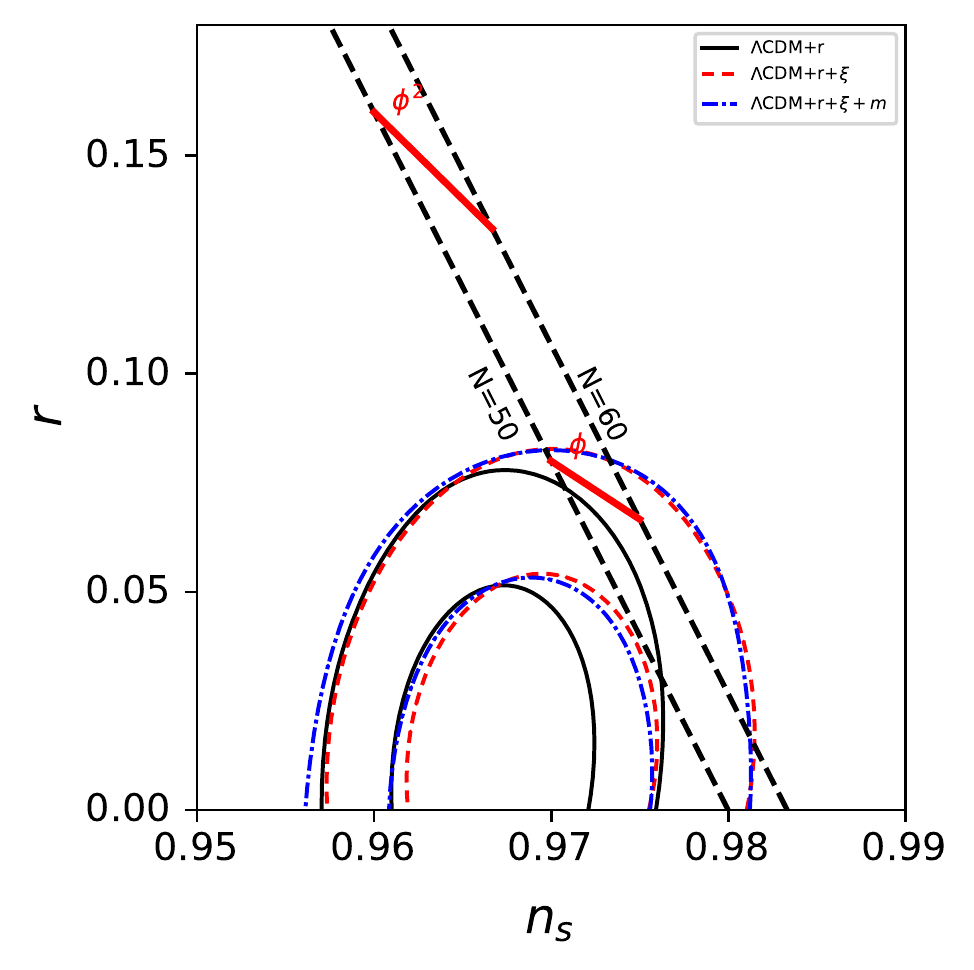}
\caption{Constraints (68\% CL and 95\% CL) in the $r-n_s$ plane by data sets \textit{Planck}(T+P)+BAO+BK15+lensing. The solid black dotted lines show the number of e-folding $N$ given by chaotic inflation model with power law potential $V \propto (\phi/M_\mathrm{Pl})^p$, while the solid red lines show allowed parameter ranges of linear and quadratic potentials between $50 < N < 60 $. The details are given in, for example, \cite{Planck_Inflation}. It is clear that $\xi$ modestly expands the allowed parameter space.}
\label{r_n_s_2D}
\end{figure}

\subsection{$\xi$ and $n_s$}
We present the 68\% CL constraints and changes of the best-fit likelihood value $\Delta\mathrm{ln}\mathcal{L}$ (with respect to the $\Lambda\mathrm{CDM}$ model) of the fitting results in Table \ref{stat table 1}. The 1D/2D triangular constraint plots of cosmological parameters are shown in Fig. \ref{appendix_fits} in the Appendix.

In Table \ref{stat table 1}, from the $\Lambda\mathrm{CDM}+\xi+m$ model, one can see that $\xi<0.352$ at 68\% CL, which is consistent with the BBN bound \cite{Man11}. The total neutrino asymmetry $L$ is smaller than 0.165 at 95\% CL. The current data sets we used are not sensitive enough to rule out $\xi=0$. The mean of $n_s$ is 0.9704, about 0.46 $\sigma$ larger than the value in the case of $\Lambda\mathrm{CDM}$, and the 68\% CL limits are also wider. In Fig. \ref{appendix_fits}, we find that even if the \textit{Planck} polarization, BAO, DES, and lensing data are included in the fitting, the $\xi-n_s$ correlation we have studied in Sec. \ref{xi_ns_Degeneracy} still holds. It is exactly this correlation and the uncertainty in $\xi$ that finally produce a loosened constraint on $n_s$. These fitting results have a major implication that $\xi$ as a physical parameter should not be omitted in data fittings of CMB data, as it can generate uncertainties on other cosmological parameters through correlations.

\begin{table*}[t]
\centering
\begin{adjustbox}{max width=\textwidth}
\begin{tabular}{*{11}{c}}
\hline
\hline 
Model & 100$\Omega_{b}h^2$ & $\Omega_{c}h^2$ & $100\theta_s$ & $\tau$ & $\mathrm{ln}(10^{10}A_s)$ & $n_s$ & $\xi$ & $r$ \\
\hline \\                 
\input{table_data_r_new.txt}
\hline
\hline
\end{tabular}
\end{adjustbox}
\caption {Parameter constraints (mean and 68\% CL values) by data sets \textit{Planck}(T+P)+BAO+BK15+lensing. The models are $L~(\Lambda\mathrm{CDM})+r$, $L+r+\xi$, and $L+r+\xi+m$, to illustrate changes in constraints of inflationary parameters induced by a non-zero $\xi$. $L+\xi+r+m$ yields $m = 0.046_{-0.046}^{+0.010}\,\mathrm{eV}$.}
\label{stat table 2}
\end{table*}

Among the 6 parameters $(\Omega_{b}h^2, \Omega_{c}h^2, 100\theta_s,\allowbreak \tau,\allowbreak \mathrm{ln}(10^{10}A_s),\allowbreak n_s)$ of the $\Lambda\mathrm{CDM}$ model, only the probability density functions (pdfs) of $\Omega_{c}h^2$, $n_s$, and $H_0$ have modest changes with the inclusion of $\xi$. In Fig. \ref{ns_1D}, the 1D marginalized pdfs of $n_s$ for selected data sets and models are plotted to show their differences. 

As a small shift in the value of $n_s$ is known to have significant impact on inflationary cosmology, we run fittings on three more models $\Lambda\mathrm{CDM}+r$, $\Lambda\mathrm{CDM}+r+\xi$, and $\Lambda\mathrm{CDM}+r+\xi+m$ based on data sets \textit{Planck}(T+P)+BAO+BK15+lensing to study this impact, where BK15 stands for the polarization data from BICEP2/\textit{Keck} \citep{BKX}. The statistics is summarized in Table \ref{stat table 2} and Fig. \ref{r_n_s_2D}.

\subsubsection{$\Lambda\mathrm{CDM}+r+\xi$}
At 68\% CL, this model gives $\xi < 1.02$, $n_s = 0.9692^{+0.0040}_{-0.0048}$, and $r < 0.0371$, compared with $n_s = 0.9670\pm 0.0036$ and $r<0.0355$ of $\Lambda\mathrm{CDM}+r$. The corresponding total neutrino asymmetry $L$ is smaller than 0.282. However, in Fig. \ref{r_n_s_2D} one can see that this model in fact expands the allowed $r-n_s$ space considerably, and it is enough to revive some inflation models excluded in the standard $\Lambda\mathrm{CDM}+r$ fitting at the 95\% CL, for example the linear $\phi$ model. Again, this is because the $\xi-n_s$ correlation we have revealed in Sec. \ref{xi_ns_Degeneracy} projects the uncertainties of $\xi$ to the $n_s$ space.

\subsubsection{$\Lambda\mathrm{CDM}+r+\xi+m$}
Further including $m$ in the fitting yields $\xi<0.424$, $n_s = 0.9692^{+0.0039}_{-0.0050}$, and $r<0.0355$ at 68\% CL. The corresponding total neutrino asymmetry $L$ is smaller than 0.108. The constraints of this model are similar to those of $\Lambda\mathrm{CDM}+r+\xi$.

\subsection{$\xi$ and $H_0$}
\begin{figure}
\centering
\includegraphics[width=10cm]{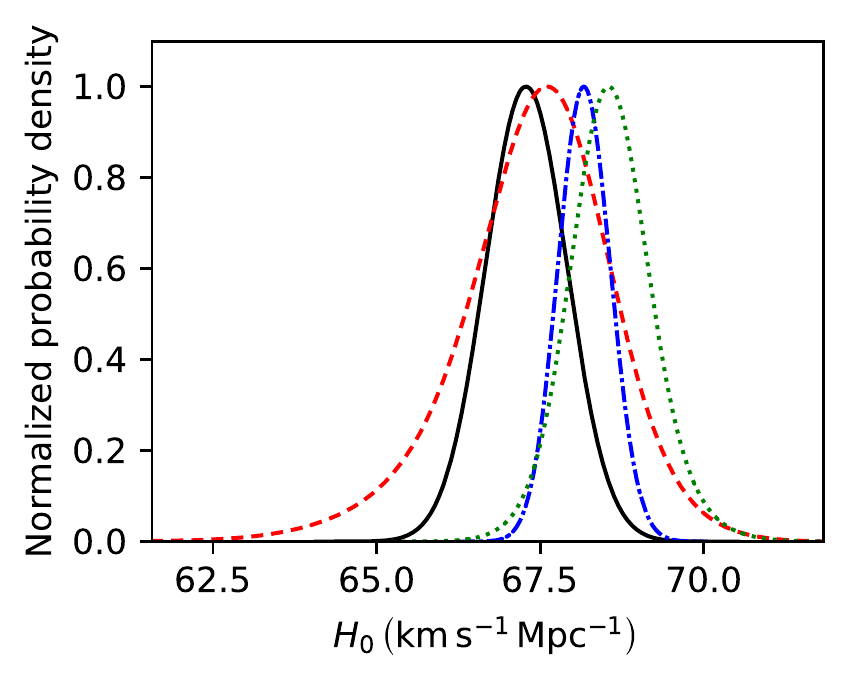}
\caption{Same as Fig. 7, but for $H_0$.}
\label{H0_1D}
\end{figure}

As the Hubble parameter $H_0$ has attracted considerable attention in recent years due to the tension in its value between cosmological surveys \citep{Planck_2018_VI} and ``local measurements'' \citep{HST_principle, Riess_2019, H0LiCOW_XIII}, we also look into the impact of $\xi$ on the constraint of $H_0$. Thus the 1D marginalized pdfs of $H_0$ are plotted in Fig. \ref{H0_1D}. 

In order to quantify the tension of $H_0$ between our fitting results and the local measurements, we apply measure 3 of \cite{sigmah}. This method can be used to quantify the statistical discordance between two pdfs and express it in terms of an equivalent Gaussian $\sigma$. We first choose HST's result to represent the typical $H_0$ constraint given by local measurements, and thus we take a Gaussian function $f(H_0)$ with mean 74.03 and $\sigma=1.42$ from \cite{Riess_2019}, to be the pdf of local measurements. On the other hand, there are pdfs of $H_0$ coming from our \texttt{CosmoMC} fittings, and so for each of these pdfs, we calculate its $\sigma_h$ with respect to $f(H_0)$. They are listed in Table \ref{stat table 1} as well.

From the $\Lambda\mathrm{CDM}+\xi+m$ model, the mean of $H_0$ is $68.57 \ {\rm km\, s^{-1}\, Mpc^{-1}}$, about 1$\sigma$ higher than the $\Lambda\mathrm{CDM}$ model. The 68\% CL limits are also wider. This is due to the correlation between $\xi$ and $H_0$ described in Sec. \ref{xi_H0_Degeneracy}. The tension between local and CMB measurements of $H_0$ is slightly reduced from $4.0\sigma$ to $3.5\sigma$ when $\xi$ is included as a parameter. 

\subsection{$\xi$ and $N_\mathrm{eff}$}
\label{section:xi_vs_neff}

\begin{figure*}
\centering
\includegraphics[width=\textwidth]{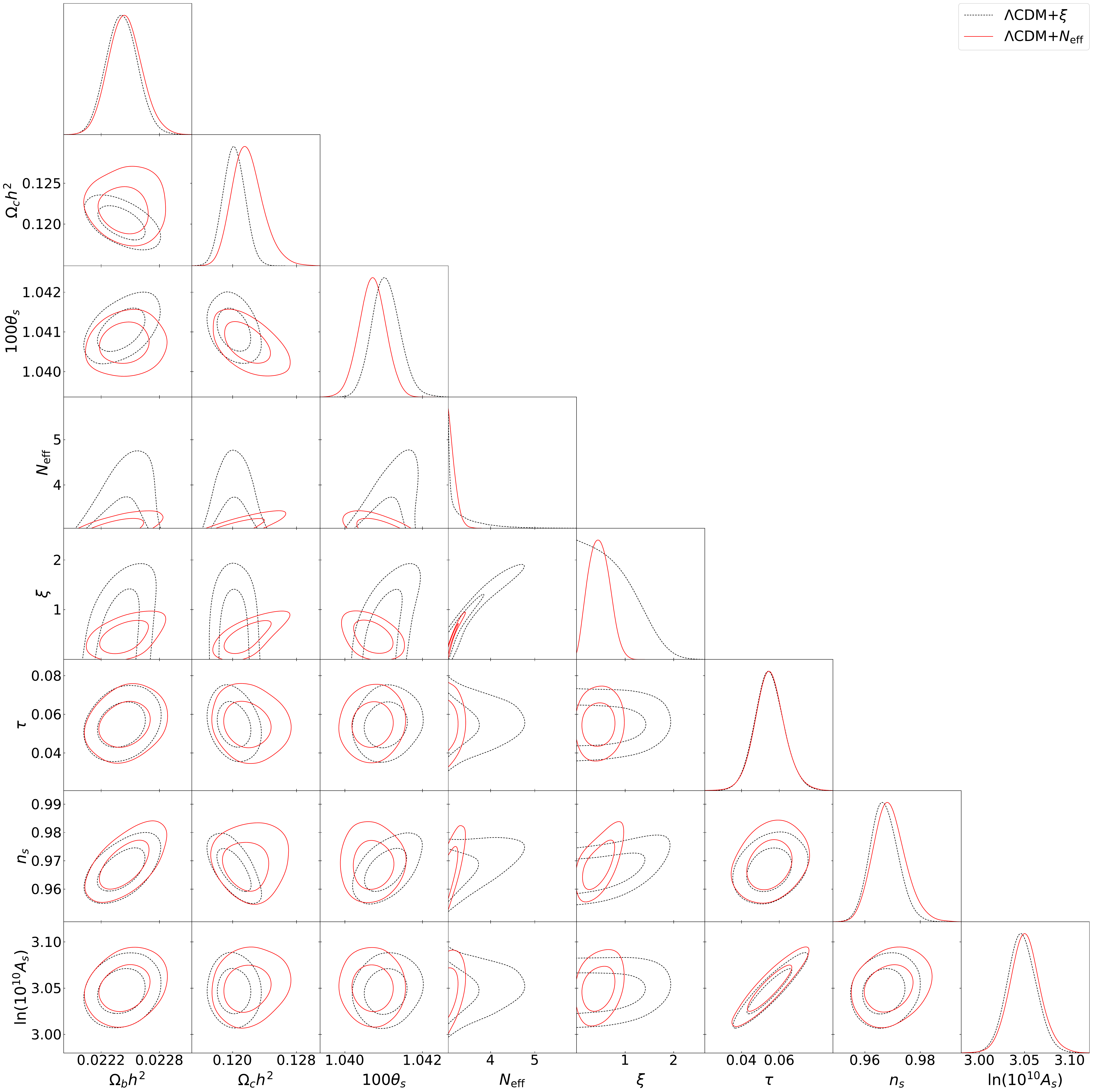}
\caption{1D marginalized posterior pdfs and 2D contours (for 68\% CL and 95\% CL) of selected cosmological parameters. The data sets involved in this fitting are \textit{Planck}(T+P)+BAO+DES+lensing, and the models are $\Lambda\mathrm{CDM}+\xi$ (black dashed lines) and $\Lambda\mathrm{CDM}+N_\mathrm{eff}$ (red solid lines). The pdfs of $\xi$ in the model $\Lambda\mathrm{CDM}+N_\mathrm{eff}$ and $N_\mathrm{eff}$ in the model $\Lambda\mathrm{CDM}+\xi$ are obtained by using Eq.~\ref{eq:neff_massless}.}
\label{xi_neff_fits}
\end{figure*}

In Fig. \ref{xi_neff_fits}, we present a direct comparison between $\xi$ vs. $N_{\mathrm{eff}}$. They are the 1D/2D triangular constraint plots on cosmological parameters based on the data set \textit{Planck}(T+P)+BAO+DES+
lensing with models $\Lambda\mathrm{CDM}+\xi$ vs. $\Lambda\mathrm{CDM}+N_\mathrm{eff}$. 
Among the six independent cosmological parameters in $\Lambda$CDM, $\Omega_{b}h^2$, $\tau$, $\mathrm{ln}(10^{10}A_s)$ and $n_s$ are almost the same in the two models, but the mean value of $\Omega_{c}h^2$ ($100\theta_s$) in the model $\Lambda\mathrm{CDM}+N_\mathrm{eff}$ is about 1$\sigma$ larger (smaller) than that of $\Lambda\mathrm{CDM}+\xi$. The 68\% CL ranges for both $\Omega_ch^2$ and $H_0$ are increased by at least 50\% in the $\Lambda\mathrm{CDM}+N_\mathrm{eff}$ model compared to those of $\Lambda\mathrm{CDM}+\xi$. This shows that $N_\mathrm{eff}$ alone cannot capture all the changes introduced by a non-zero $\xi$, given the current set of cosmological data. In the model $\Lambda\mathrm{CDM}+\xi$, the value of $N_\mathrm{eff}$ is fixed to 3.046. However, in Fig.~\ref{xi_neff_fits}, we add the equivalent pdfs of $N_\mathrm{eff}$ by using Eq.~\ref{eq:neff_massless} to transform the pdfs of $\xi$, under the assumption that neutrinos are massless. The pdfs of $\xi$ in the model $\Lambda\mathrm{CDM}+N_\mathrm{eff}$ are obtained similarly.





\section{Conclusions}
\label{Conclusions}
In this paper we study the implications of a non-zero neutrino asymmetry on cosmological parameters extracted from cosmological data. We have calculated the cosmological neutrino energy density for general $\xi$ with the mixing of neutrino flavor eigenstates taken into account. We reveal the correlation between $\xi$ and $H_0$, and $\xi$ and $n_s$ in CMB fittings, and we explain their origins using semi-analytical methods. The current data allow a range of $\xi \lesssim 1$, and there are additional uncertainties in other cosmological parameters extracted from data fitting when $\xi$ is included as a parameter. Therefore, $\xi$ can be considered to be a non-negligible systematic in the constraints of the $r-n_s$ space and $H_0$.

The reason that a non-zero $\xi$ alleviates the tension on $H_0$ is similar to that of a sterile neutrino \citep{Hu1, Hu2}. Both essentially increase the neutrino energy density, so that degeneracies between these parameters and $H_0$ like the one shown in Sec. \ref{xi_H0_Degeneracy} are induced. However, we emphasize that $\xi$ is a standard physical parameter. A non-zero $\xi$ does not require any new physics, unlike the existence of sterile neutrinos. Moreover, we further demonstrate that including $\xi$ as a parameter loosen the constraint on $n_s$, which is critical for selection of inflation models. 

The true values of relic neutrino degeneracies are important and interesting themselves. Future observations such as the high precision measurement of matter power spectrum will help to measure or tighten the constraints on $\xi$ \citep{Carton}. This will also reduce a systematic uncertainty on cosmological parameters determined from CMB measurement.

\section{Acknowledgements}
We thank Clem Pryke and Keith Olive for helpful discussion. This work is partially supported by grants from the Research Grant Council of the Hong Kong Special Administrative Region, China (Project No.s AoE/P-404/18, 14301214) and the VC Discretionary Fund of The Chinese University of Hong Kong. We acknowledge the support of the CUHK Central Research Cluster, on which parts of the computation in this work have been performed. 

\appendix
\section{Results of \texttt{CosmoMC} fittings}

In Fig. \ref{appendix_fits} and Table \ref{stat table 3}, we present 68\% CL constraints as well as $\Delta\mathrm{ln}\mathcal{L}$ of fitting results of all combinations of models and data sets mentioned in Sec \ref{Results}. We elaborate them model by model in the subsequent subsections. Hereafter all $H_0$ values are presented in the unit of ${\rm km\, s^{-1}\, Mpc^{-1}}$.

\subsection{$\Lambda\mathrm{CDM}+\xi$}

\begin{figure*}
\centering
\includegraphics[width=\textwidth]{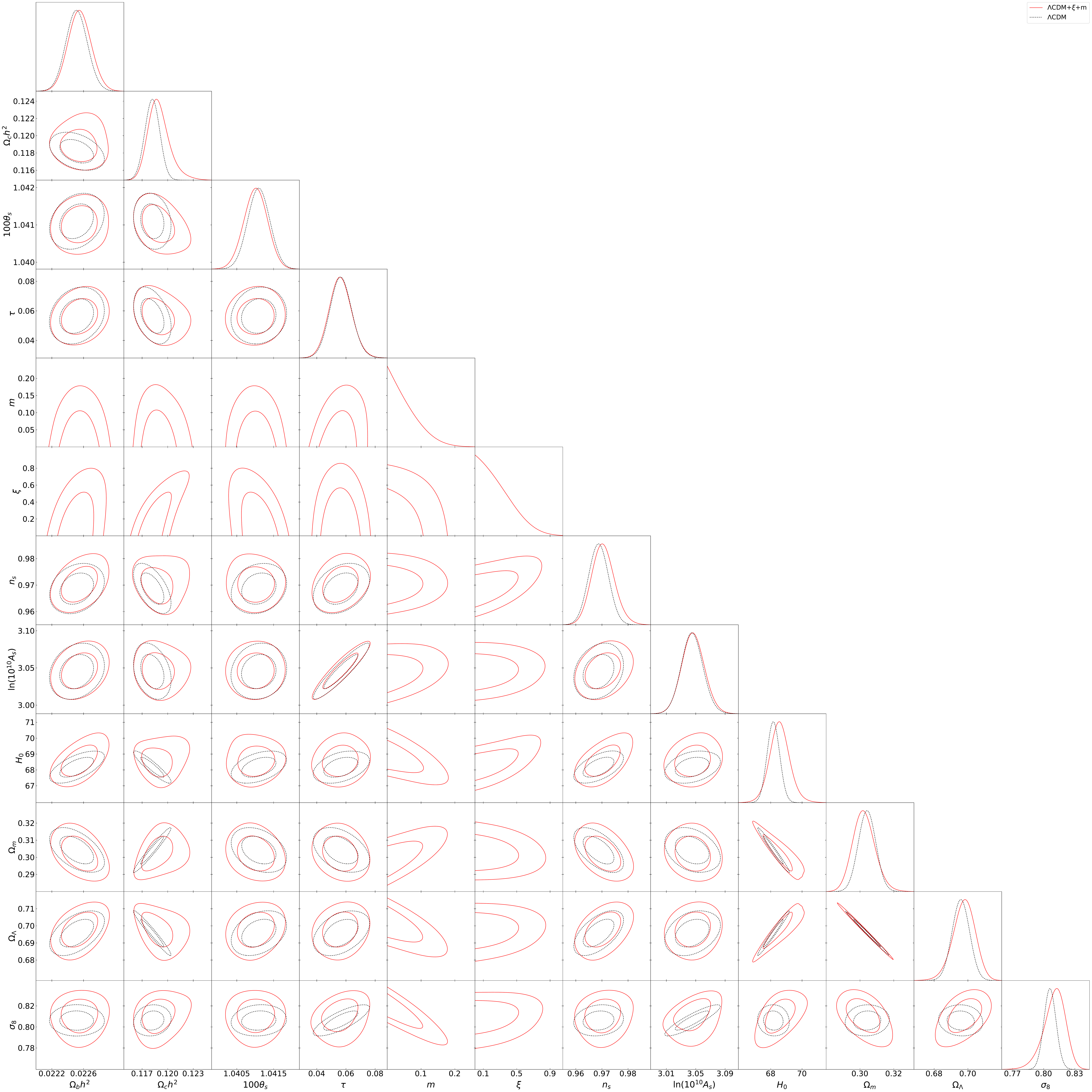}
\caption{1D marginalized posterior pdfs and 2D contours (for 68\% CL and 95\% CL) of cosmological parameters. The data sets involved in this fitting are \textit{Planck}(T+P)+BAO+DES+lensing, and the models are $\Lambda\mathrm{CDM}$ (black dashed lines) and $\Lambda\mathrm{CDM}+\xi+m$ (red solid lines).}
\label{appendix_fits}
\end{figure*}

Under data set \textit{Planck}(T), the mean of $\xi$ is $2.26$, and the tight constraints on $n_s$ and $H_0$ in $\Lambda\mathrm{CDM}$ are loosened with the inclusion of $\xi$ as a cosmological parameter, because of the extra degeneracy available. Both $n_s$ and $H_0$ tend to increase. However, in the meantime $\tau$ also drifts to a large value $\tau = 0.161$, since the temperature data is inadequate to fix it. As indicated in Fig. \ref{appendix_fits}, there are $n_s-\tau$ and $H_0-\tau$ correlations that allow $n_s$ and $H_0$ to increase, given that there is space for $\tau$ variation. Thus, when \textit{Planck}(T+P) is used, $\tau$, $n_s$, and $H_0$ decrease to values similar to those of the $\Lambda\mathrm{CDM}$ case. Also, once we add the \textit{Planck} polarization data, the constraint of $\xi$ is almost the same (with mean $\approx 0.7$) no matter what additional data sets we consider.

Nevertheless, it should be noted that, all the 6 major cosmological parameters do not show any significant changes even with the addition of $\xi$. They are still compatible with the standard values from \textit{Planck}(T+P) \& $\Lambda\mathrm{CDM}$.

\begin{table*}
\centering
\begin{adjustbox}{max width=\textwidth}
\begin{tabular}{*{14}{c}}
\hline
\hline 
Data Set & Model & $\Delta\mathrm{ln}\mathcal{L}$ & $\sigma_h$ & $H_0$ & 100$\Omega_{b}h^2$ & $\Omega_{c}h^2$ & $100\theta_s$ & $\tau$ & $\mathrm{ln}(10^{10}A_s)$ & $n_s$ & $\xi$ & $N_{\mathrm{eff}}$ & $m$ \\
\hline \\

\input{table_data_sigma_nnu.txt}
\hline
\hline
\end{tabular}
\end{adjustbox} 
\caption {Parameter constraints (mean and 68\% CL values), $\Delta\mathrm{ln}\mathcal{L}$ and $\sigma_h$ from various combinations of models and data sets. Note that \textit{P}(T) and \textit{P}(T+P) stands for \textit{Planck} Temperature and \textit{Planck} Temperature+Polarization respectively. $L$ represents the $\Lambda\mathrm{CDM}$ case, which is the base-line model with 6 parameters $\{\Omega_{b}h^2, \Omega_{c}h^2, 100\theta_s, \tau, \mathrm{ln}(10^{10}A_s), n_s\}$. $\xi = |\xi_{\nu_3}|$ and $m$ is the sum of neutrino mass eigenvalues in eV.}
\label{stat table 3}
\end{table*}

\subsection{$\Lambda\mathrm{CDM}+\xi+m$}
\label{LCDM_xi_m}
As the neutrino mass is now well-known to be non-zero, we include the total neutrino mass $m$ as a variable as well. In this model, the constraint of $\xi$ changes a bit for different combinations of the data sets. Excluding the case \textit{Planck}(T), \textit{Planck}(T+P)+BAO+DES+lensing gives the tightest constraint of $\xi<0.352$, while \textit{Planck}(T+P)+BAO gives the widest constraint of $\xi<0.435$. Fitting to the data sets \textit{Planck}(T+P), at 68\% CL $H_0$ changes from $67.41^{+0.62}_{-0.61}$ to $67.43^{+1.21}_{-0.91}$ while $n_s$ stays almost the same, compared with the $\Lambda\mathrm{CDM}+\xi$ case. Replacing the data sets by \textit{Planck}(T+P)+BAO, $H_0$ changes from $67.77^{+0.46}_{-0.47}$ to $68.18^{+0.53}_{-0.73}$ while $n_s$ barely changes, with $m<0.059$. The BAO data is able to put a better constraint on the neutrino mass. 

\subsection{$\Lambda\mathrm{CDM}+N_{\mathrm{eff}}+m$}

The parameter constraints are almost the same after adding neutrino mass to the model $\Lambda\mathrm{CDM}+N_{\mathrm{eff}}$. Similar to the comparison between $\Lambda\mathrm{CDM}+\xi$ and $\Lambda\mathrm{CDM}+N_{\mathrm{eff}}$, in this model the values of $n_s$ and $H_0$ are slightly larger than the model in Section \ref{LCDM_xi_m}.

\bibliography{database}{}
\bibliographystyle{JHEP}

\end{document}

%% file: table_data_1_new.txt
$L$
& 0.00 
& 4.0 
& $68.17^{+0.38}_{-0.39}$
& $2.251\pm 0.013$
& $0.11820^{+0.00084}_{-0.00085}$
& $1.04110\pm 0.00029$
& $0.0566\pm 0.0072$
& $3.045\pm 0.014$
& $0.9687\pm 0.0037$
& ---
& ---\\
$L+\xi+m$
& 0.25 
& 3.5 
& $68.57^{+0.56}_{-0.62}$
& $2.255^{+0.013}_{-0.015}$
& $0.11896^{+0.00091}_{-0.00139}$
& $1.04101\pm 0.00031$
& $0.0562^{+0.0070}_{-0.0077}$
& $3.046^{+0.014}_{-0.015}$
& $0.9704^{+0.0038}_{-0.0044}$
& $0.274^{+0.078}_{-0.274}$
& $0.055^{+0.013}_{-0.055}$\\\\

%% file: table_data_r_new.txt
$L+r$
& $2.241\pm 0.013$
& $0.11936^{+0.00094}_{-0.00093}$
& $1.04100\pm 0.00029$
& $0.0570^{+0.0066}_{-0.0074}$
& $3.049^{+0.013}_{-0.014}$
& $0.9670\pm 0.0036$
& ---
& $0.0281^{+0.0074}_{-0.0276}$\\
$L+r+\xi$
& $2.247\pm 0.014$
& $0.11942\pm 0.00093$
& $1.04117^{+0.00031}_{-0.00038}$
& $0.0572^{+0.0068}_{-0.0075}$
& $3.051\pm 0.014$
& $0.9692^{+0.0040}_{-0.0048}$
& $0.78^{+0.24}_{-0.78}$
& $0.0288^{+0.0083}_{-0.0271}$\\
$L+r+\xi+m$
& $2.247^{+0.014}_{-0.015}$
& $0.1205^{+0.0011}_{-0.0017}$
& $1.04087\pm 0.00032$
& $0.0561^{+0.0067}_{-0.0077}$
& $3.050^{+0.014}_{-0.016}$
& $0.9692^{+0.0039}_{-0.0050}$
& $0.327^{+0.097}_{-0.327}$
& $0.0280^{+0.0075}_{-0.0271}$\\
\\

%% file: table_data_sigma_nnu.txt
\multirow{5}{3em}{\textit{P}(T)}
&$L$
& 0.00 
& 3.1 
& $68.1\pm 1.2$
& $2.238\pm 0.028$
& $0.1180\pm 0.0026$
& $1.04112^{+0.00052}_{-0.00053}$
& $0.108\pm 0.033$
& $3.145^{+0.065}_{-0.060}$
& $0.9714^{+0.0078}_{-0.0079}$
& ---
& ---
& ---\\
&$L+\xi$
& 1.58 
& 1.2 
& $70.9^{+1.8}_{-3.1}$
& $2.308^{+0.044}_{-0.074}$
& $0.1150\pm 0.0034$
& $1.0428^{+0.0010}_{-0.0017}$
& $0.161^{+0.047}_{-0.058}$
& $3.254^{+0.090}_{-0.116}$
& $1.001^{+0.015}_{-0.031}$
& $2.26^{+1.26}_{-0.77}$
& ---
& ---\\
&$L+\xi+m$
& 0.95 
& -0.0 
& $74.5^{+5.8}_{-10.6}$
& $2.336^{+0.060}_{-0.117}$
& $0.1244^{+0.0041}_{-0.0057}$
& $1.04064\pm 0.00058$
& $0.206^{+0.058}_{-0.084}$
& $3.35^{+0.12}_{-0.17}$
& $1.021^{+0.023}_{-0.055}$
& $1.22^{+0.70}_{-0.50}$
& ---
& $0.331^{+0.057}_{-0.331}$\\
&$L+N_{\mathrm{eff}}$
& 1.81 
& -0.6 
& $82.1^{+5.6}_{-10.3}$
& $2.404^{+0.077}_{-0.116}$
& $0.1271^{+0.0046}_{-0.0065}$
& $1.04072\pm 0.00052$
& $0.225^{+0.065}_{-0.074}$
& $3.40^{+0.13}_{-0.15}$
& $1.046^{+0.033}_{-0.051}$
& ---
& $4.42^{+0.53}_{-1.02}$
& ---\\
&$L+N_{\mathrm{eff}}+m$
& 0.97 
& -0.2 
& $78.5^{+7.0}_{-10.2}$
& $2.388^{+0.079}_{-0.110}$
& $0.1277^{+0.0048}_{-0.0065}$
& $1.04045\pm 0.00058$
& $0.238^{+0.067}_{-0.068}$
& $3.42\pm 0.14$
& $1.043^{+0.034}_{-0.048}$
& ---
& $4.40^{+0.56}_{-0.98}$
& $0.406^{+0.083}_{-0.406}$\\
\\\multirow{5}{3em}{\textit{P}(T+P)}
&$L$
& 0.00 
& 4.4 
& $67.29^{+0.60}_{-0.61}$
& $2.236\pm 0.015$
& $0.1202^{+0.0013}_{-0.0014}$
& $1.04089^{+0.00032}_{-0.00031}$
& $0.0544^{+0.0072}_{-0.0080}$
& $3.045^{+0.015}_{-0.016}$
& $0.9649\pm 0.0044$
& ---
& ---
& ---\\
&$L+\xi$
& -0.52 
& 4.3 
& $67.41^{+0.62}_{-0.61}$
& $2.241\pm 0.016$
& $0.1202\pm 0.0014$
& $1.04106^{+0.00033}_{-0.00038}$
& $0.0548^{+0.0073}_{-0.0080}$
& $3.047\pm 0.016$
& $0.9670^{+0.0047}_{-0.0053}$
& $0.73^{+0.22}_{-0.73}$
& ---
& ---\\
&$L+\xi+m$
& 0.26 
& 3.6 
& $67.43^{+1.21}_{-0.91}$
& $2.240\pm 0.016$
& $0.1211^{+0.0015}_{-0.0018}$
& $1.04079^{+0.00033}_{-0.00034}$
& $0.0546^{+0.0073}_{-0.0080}$
& $3.048\pm 0.016$
& $0.9669^{+0.0046}_{-0.0054}$
& $0.300^{+0.086}_{-0.300}$
& ---
& $0.091^{+0.014}_{-0.091}$\\
&$L+N_{\mathrm{eff}}$
& -1.36 
& 3.4 
& $68.03^{+0.72}_{-1.00}$
& $2.245^{+0.016}_{-0.018}$
& $0.1219^{+0.0016}_{-0.0022}$
& $1.04071\pm 0.00034$
& $0.0549^{+0.0073}_{-0.0080}$
& $3.051\pm 0.016$
& $0.9689^{+0.0051}_{-0.0062}$
& ---
& $3.162^{+0.026}_{-0.116}$
& ---\\
&$L+N_{\mathrm{eff}}+m$
& -0.88 
& 3.3 
& $67.8^{+1.3}_{-1.0}$
& $2.245^{+0.016}_{-0.018}$
& $0.1219^{+0.0016}_{-0.0021}$
& $1.04070\pm 0.00034$
& $0.0551^{+0.0072}_{-0.0079}$
& $3.050^{+0.015}_{-0.017}$
& $0.9689^{+0.0051}_{-0.0063}$
& ---
& $3.160^{+0.024}_{-0.114}$
& $0.091^{+0.014}_{-0.091}$\\
\\\multirow{5}{3em}{\textit{P}(T+P)\\+BAO}
&$L$
& 0.00 
& 4.3 
& $67.66\pm 0.45$
& $2.242^{+0.013}_{-0.014}$
& $0.1193\pm 0.0010$
& $1.04100\pm 0.00029$
& $0.0554^{+0.0076}_{-0.0077}$
& $3.045^{+0.015}_{-0.017}$
& $0.9668^{+0.0038}_{-0.0037}$
& ---
& ---
& ---\\
&$L+\xi$
& -0.15 
& 4.2 
& $67.77^{+0.46}_{-0.47}$
& $2.247^{+0.015}_{-0.014}$
& $0.1194\pm 0.0010$
& $1.04117^{+0.00031}_{-0.00038}$
& $0.0560\pm 0.0076$
& $3.048\pm 0.016$
& $0.9690^{+0.0041}_{-0.0047}$
& $0.76^{+0.23}_{-0.76}$
& ---
& ---\\
&$L+\xi+m$
& 0.58 
& 3.6 
& $68.18^{+0.53}_{-0.73}$
& $2.247^{+0.014}_{-0.015}$
& $0.1205^{+0.0011}_{-0.0018}$
& $1.04090^{+0.00035}_{-0.00031}$
& $0.0556^{+0.0071}_{-0.0079}$
& $3.049^{+0.015}_{-0.017}$
& $0.9689^{+0.0040}_{-0.0050}$
& $0.342^{+0.093}_{-0.342}$
& ---
& $0.048^{+0.011}_{-0.048}$\\
&$L+N_{\mathrm{eff}}$
& -0.42 
& 3.4 
& $68.43^{+0.59}_{-0.83}$
& $2.251\pm 0.015$
& $0.1213^{+0.0014}_{-0.0022}$
& $1.04080^{+0.00036}_{-0.00032}$
& $0.0561^{+0.0074}_{-0.0081}$
& $3.051^{+0.016}_{-0.017}$
& $0.9711^{+0.0045}_{-0.0054}$
& ---
& $3.170^{+0.029}_{-0.124}$
& ---\\
&$L+N_{\mathrm{eff}}+m$
& -0.21 
& 3.3 
& $68.52^{+0.65}_{-0.86}$
& $2.251\pm 0.015$
& $0.1214^{+0.0014}_{-0.0022}$
& $1.04078\pm 0.00033$
& $0.0557\pm 0.0076$
& $3.051^{+0.016}_{-0.017}$
& $0.9709^{+0.0045}_{-0.0055}$
& ---
& $3.172^{+0.030}_{-0.126}$
& $0.052^{+0.012}_{-0.052}$\\
\\\multirow{5}{3em}{\textit{P}(T+P)\\+BAO\\+DES}
&$L$
& 0.00 
& 3.9 
& $68.24^{+0.40}_{-0.41}$
& $2.251\pm 0.013$
& $0.11801^{+0.00088}_{-0.00089}$
& $1.04111\pm 0.00029$
& $0.0541^{+0.0075}_{-0.0074}$
& $3.039^{+0.016}_{-0.015}$
& $0.9692\pm 0.0037$
& ---
& ---
& ---\\
&$L+\xi$
& 0.56 
& 3.8 
& $68.34\pm 0.42$
& $2.256^{+0.014}_{-0.015}$
& $0.11806\pm 0.00087$
& $1.04126^{+0.00032}_{-0.00035}$
& $0.0545^{+0.0076}_{-0.0075}$
& $3.041\pm 0.016$
& $0.9712^{+0.0039}_{-0.0046}$
& $0.73^{+0.22}_{-0.73}$
& ---
& ---\\
&$L+\xi+m$
& -1.01 
& 3.4 
& $68.54^{+0.64}_{-0.66}$
& $2.256\pm 0.014$
& $0.11882^{+0.00095}_{-0.00149}$
& $1.04101\pm 0.00031$
& $0.0545^{+0.0075}_{-0.0076}$
& $3.042\pm 0.016$
& $0.9713^{+0.0039}_{-0.0046}$
& $0.289^{+0.082}_{-0.289}$
& ---
& $0.073^{+0.017}_{-0.073}$\\
&$L+N_{\mathrm{eff}}$
& -0.54 
& 3.2 
& $68.87^{+0.48}_{-0.73}$
& $2.259\pm 0.014$
& $0.1195^{+0.0011}_{-0.0018}$
& $1.04095^{+0.00033}_{-0.00031}$
& $0.0545^{+0.0076}_{-0.0075}$
& $3.044\pm 0.016$
& $0.9726^{+0.0041}_{-0.0049}$
& ---
& $3.146^{+0.022}_{-0.100}$
& ---\\
&$L+N_{\mathrm{eff}}+m$
& -0.57 
& 3.2 
& $68.84^{+0.68}_{-0.79}$
& $2.259^{+0.014}_{-0.015}$
& $0.1196^{+0.0012}_{-0.0019}$
& $1.04093\pm 0.00032$
& $0.0548\pm 0.0077$
& $3.044\pm 0.016$
& $0.9729^{+0.0042}_{-0.0052}$
& ---
& $3.152^{+0.024}_{-0.106}$
& $0.076^{+0.018}_{-0.076}$\\
\\\multirow{5}{3em}{\textit{P}(T+P)\\+BAO\\+DES\\+lensing}
&$L$
& 0.00 
& 4.0 
& $68.17^{+0.38}_{-0.39}$
& $2.251\pm 0.013$
& $0.11820^{+0.00084}_{-0.00085}$
& $1.04110\pm 0.00029$
& $0.0566\pm 0.0072$
& $3.045\pm 0.014$
& $0.9687\pm 0.0037$
& ---
& ---
& ---\\
&$L+\xi$
& 0.72 
& 3.9 
& $68.27\pm 0.40$
& $2.256\pm 0.014$
& $0.11821^{+0.00085}_{-0.00084}$
& $1.04125^{+0.00030}_{-0.00035}$
& $0.0569^{+0.0069}_{-0.0076}$
& $3.047^{+0.014}_{-0.015}$
& $0.9706^{+0.0038}_{-0.0045}$
& $0.71^{+0.21}_{-0.71}$
& ---
& ---\\
&$L+\xi+m$
& 0.25 
& 3.5 
& $68.57^{+0.56}_{-0.62}$
& $2.255^{+0.013}_{-0.015}$
& $0.11896^{+0.00091}_{-0.00139}$
& $1.04101\pm 0.00031$
& $0.0562^{+0.0070}_{-0.0077}$
& $3.046^{+0.014}_{-0.015}$
& $0.9704^{+0.0038}_{-0.0044}$
& $0.274^{+0.078}_{-0.274}$
& ---
& $0.055^{+0.013}_{-0.055}$\\
&$L+N_{\mathrm{eff}}$
& 1.56 
& 3.3 
& $68.79^{+0.47}_{-0.74}$
& $2.259^{+0.014}_{-0.015}$
& $0.1197^{+0.0011}_{-0.0017}$
& $1.04093^{+0.00031}_{-0.00032}$
& $0.0568^{+0.0068}_{-0.0075}$
& $3.049\pm 0.014$
& $0.9721^{+0.0041}_{-0.0048}$
& ---
& $3.143^{+0.021}_{-0.097}$
& ---\\
&$L+N_{\mathrm{eff}}+m$
& 1.20 
& 3.2 
& $68.84^{+0.62}_{-0.73}$
& $2.259\pm 0.014$
& $0.1196^{+0.0011}_{-0.0017}$
& $1.04093\pm 0.00032$
& $0.0565^{+0.0070}_{-0.0078}$
& $3.049^{+0.014}_{-0.016}$
& $0.9721^{+0.0040}_{-0.0049}$
& ---
& $3.141^{+0.019}_{-0.095}$
& $0.061^{+0.014}_{-0.061}$\\
\\